\documentclass[prd,aps,twocolumn,preprintnumbers, showpacs, nofootinbib,superscriptaddress,notitlepage]{revtex4-1}
\usepackage{stmaryrd}
\usepackage{epsfig}
\usepackage{amsfonts}
\usepackage{graphicx}
\usepackage{color}
\usepackage{amssymb,amsmath,psfrag,slashed,graphicx}
\usepackage[normalem]{ulem}
\usepackage[utf8]{inputenc}

\begin{document}

\preprint{MIT-CTP/5073}

\title{Matching the Quasi Meson Distribution Amplitude in RI/MOM scheme}

\author{Yu-Sheng Liu}
\email{mestelqure@gmail.com}
\affiliation{Tsung-Dao Lee Institute, Shanghai Jiao-Tong University, Shanghai 200240, China}

\author{Wei Wang}
\email{wei.wang@sjtu.edu.cn}
\affiliation{SKLPPC, School of Physics and Astronomy, Shanghai Jiao-Tong University, Shanghai 200240, China}

\author{Ji Xu}
\email{xuji1991@sjtu.edu.cn}
\affiliation{SKLPPC, School of Physics and Astronomy, Shanghai Jiao-Tong University, Shanghai 200240, China}

\author{Qi-An Zhang}
\email{zhangqa@ihep.ac.cn}
\affiliation{Institute of High Energy Physics, Chinese Academy of Science, Beijing 100049, China}
\affiliation{School of Physics, University of Chinese Academy of Sciences, Beijing 100049, China}

\author{Shuai Zhao}
\email{shuai.zhao@sjtu.edu.cn}
\affiliation{SKLPPC, School of Physics and Astronomy, Shanghai Jiao-Tong University, Shanghai 200240, China}

\author{Yong Zhao}
\email{yzhaoqcd@mit.edu}
\affiliation{Center for Theoretical Physics, Massachusetts Institute of Technology, Cambridge, Massachusetts 02139, USA}

\begin{abstract}
The $x$-dependence of light-cone distribution amplitude (LCDA) can be directly calculated from a quasi distribution amplitude (DA) in lattice QCD within the framework of large-momentum effective theory (LaMET). In this paper, we study the one-loop renormalization of the quasi-DA in the regularization-independent momentum subtraction (RI/MOM) scheme. The renormalization factor for the quasi parton distribution function can be used to renormalize the quasi-DA provided that they are implemented on lattice and in perturbation theory in the same manner. We derive the one-loop matching coefficient that matches quasi-DA in the RI/MOM scheme onto LCDA in the $\overline{\rm MS}$ scheme. Our result provides the crucial step to extract the LCDAs from lattice matrix elements of quasi-DAs.
\end{abstract}

\maketitle

\section{Introduction}

Exclusive processes at high energy play a vital role in understanding  the strong interactions in hadronic reactions. The large momentum transfer in many processes guarantees the use of operator product expansion that  separates   short-distance and long-distance degrees of freedom.   This separation is often achieved through the factorization theorem. When collinear factorization is applicable, the scattering/decay amplitude of a hard exclusive reaction can then be written in terms of a convolution of a hard-scattering kernel with a nonperturbative function---light-cone distribution amplitudes (LCDAs). The LCDAs characterize the momentum distribution of quarks and anti-quarks inside a meson.

LCDAs show a few facets. First, in the perturbative region, the evolution of  LCDAs is  governed by the renormalization group equation,  namely Efremov-Radyushkin-Brodsky-Lepage (ERBL) equation~\cite{Efremov:1978rn,Efremov:1979qk,Lepage:1979zb,Lepage:1980fj}. Large logarithms of  hard momentum scale  and   hadronic scale in the amplitude  can be resummed using renormalization group.  Secondly, LCDAs can be expanded into series of Gegenbauer polynomials which are eigenfunctions of the ERBL kernel. When the involved energy is high, we expect that LCDA in pion or kaon approaches its asymptotic form,  $6x(1-x)$, for the leading-twist contributions.  At accessible energies,  such an expectation is challenged,  for instance,  by the scaling violation in  the BaBar measurement~\cite{Aubert:2009mc} of  the   $\gamma\gamma^*\to \pi$ form factor. Higher Gegenbauer moments  and high power corrections  are shown not  negligible~\cite{Wang:2017ijn}.   Most notably,  being non-perturbative in nature, LCDAs cannot be evaluated in perturbation theory. Our current knowledge on   LCDAs    largely relies  on  phenomenological approaches like QCD sum rules or global analyses of data.  Non-perturbative approaches based on first principle methods, e.g., lattice QCD (LQCD) can be utilized  to calculate only the lowest moments of LCDAs.  So far, the pion LCDA is only known up to its second moment from  LQCD~\cite{Braun:2015axa,Bali:2017ude}. The kaon LCDA, an indispensable input to calculating the differential decay width of $B\to K\ell^+\ell^-$ and probing new physics effects therein~\cite{Li:2018lxi}, also receives light efforts from the lattice~\cite{Braun:2006dg,Boyle:2006pw,Arthur:2010xf}.

Recently, a groundbreaking approach to calcualte LCDA and more general parton physics from lattice QCD is formulated as large-momentum effective theory (LaMET)~\cite{Ji:2013dva,Ji:2014gla}, where the full $x$-dependence of LCDA's as well as other parton distributions can be accessed. In LaMET, instead of directly calculating light-cone correlations, one can start from equal-time correlations in a large-momentum hadron state, which are known as quasi parton distribution functions (quasi-PDFs), quasi-DAs, etc. At finite but large hadron momentum $P^z\gg\Lambda_{\rm QCD}$, the quasi observables can be factorized as the convolution of a perturbatively calculable matching coefficient and the corresponding light-cone observable, up to power corrections suppressed by $1/P^z$.  Through this factorization, one can extract light-cone observables from quasi ones calculated on the lattice.

In the past years, vast progresses have  been made in the development of LaMET. 
The factorization formulas have been studied for the cases of flavor-nonsinglet quasi quark PDFs~\cite{Xiong:2013bka,Ma:2014jla,Izubuchi:2018srq}, transverse momentum dependent (TMD) PDFs~\cite{Ji:2014hxa,Ji:2018hvs,Ebert:2018xxx}, generalized parton distributions (GPDs)~\cite{Ji:2015qla,Xiong:2015nua} and DAs~\cite{Xu:2018mpf}, and the gluon PDF~\cite{Wang:2017qyg,Wang:2017eel}. The effectiveness of LaMET has also been explored in solvable scenarios for QCD such as heavy quarkonia~\cite{Jia:2015pxx} and 1+1 dimensional theories~\cite{Jia:2017uul,Jia:2018qee,Ji:2018waw}. The multiplicative renormalizability of quasi-PDF in coordinate space has been proven for the quark~\cite{Ji:2015jwa,Ji:2017oey,Green:2017xeu,Ishikawa:2017faj} and gluon~\cite{Zhang:2018diq,Li:2018tpe} cases, which enables a nonperturbative renormalization of the quasi-PDFs on the lattice using the regularization independent momentum subtraction (RI/MOM) scheme~\cite{Martinelli:1994ty,Alexandrou:2017huk,Constantinou:2017sej,Chen:2017mzz,Lin:2017ani,Stewart:2017tvs,Liu:2018uuj}. Meanwhile, the lattice calculations of parton distributions with LaMET have been carried out and improved over the past years~\cite{Lin:2014zya,Alexandrou:2015rja,Chen:2016utp,Alexandrou:2016jqi,Zhang:2017bzy,Alexandrou:2017huk,Chen:2017gck,Alexandrou:2018pbm,Chen:2018xof,Chen:2018fwa,Alexandrou:2018eet,Lin:2018qky,Fan:2018dxu,Liu:2018hxv}, and the most recent results at physical pion mass~\cite{Alexandrou:2018pbm,Chen:2018xof,Alexandrou:2018eet,Lin:2018qky,Liu:2018hxv} and large nucleon momenta~\cite{Chen:2018xof,Lin:2018qky,Liu:2018hxv} have seen remarkable agreements with the global analysis of PDFs in the moderate-to-large $x$ region (see \cite{Gao:2017yyd} for a recent review).

For quasi-DAs,  the matching coefficients have been calculated in dimensional regularization and transverse-momentum cutoff schemes~\cite{Zhang:2017bzy,Bali:2017gfr,Chen:2017gck,Xu:2018mpf}. Unfortunately, neither scheme is suitable for a nonperturbative renormalization of the quasi-PDF on the lattice. The RI/MOM scheme was proposed to serve this purpose~\cite{Constantinou:2017sej,Stewart:2017tvs} and has been used for the lattice renormalization of quasi-PDFs~\cite{Alexandrou:2017huk,Chen:2017mzz,Lin:2017ani,Alexandrou:2018pbm,Chen:2018xof,Alexandrou:2018eet,Liu:2018uuj,Lin:2018qky}. Since the quasi-DAs are defined from the same spatial correlator as the quasi-PDFs, the RI/MOM scheme can be readily applied to their lattice renormalization. However, a perturbative matching coefficient that converts the quasi-DA in the RI/MOM scheme to LCDA in $\overline{\rm MS}$ scheme is still not available yet.
In this work, our main motif is to calculate this matching coefficient at one loop.
Our result will be a key element of the lattice calculation of LCDAs with LaMET.

The rest of this paper is organized as follows: In Sec.~\ref{sec:DA}, we briefly review the twist-2 LCDA and quasi-DA.  In Sec.~\ref{sec:renormalization}, we review the RI/MOM scheme on the lattice. In Sec.~\ref{sec:oneloop}, we show one-loop matching coefficients from quasi-DAs in the RI/MOM scheme to LCDAs in $\overline {\rm MS}$ scheme. A summary is presented in Sec.~\ref{sec:sum}.

\section{Distribution amplitude}\label{sec:DA}
To define the meson LCDA, we introduce the Fourier transform of a light-cone correlator $\langle P,\epsilon|O(\bar{\Gamma},\xi^-)|0\rangle$,
\begin{align}\label{eq:LCDA_numerator}
{\cal F}(\bar{\Gamma},x,\mu)=P^+\int \frac{d\xi^-}{2\pi}e^{-i x\xi^- P^+}\langle P,\epsilon|O(\bar{\Gamma},\xi^-)|0\rangle
\end{align}
where $\xi^\pm=(\xi^0\pm\xi^3)/\sqrt{2}$; $x\in[0,1]$ is the momentum fraction of quark with respect to meson in the + direction; $\mu$ is the renormalization scale; the meson state $|P,\epsilon\rangle$ is denoted by its momentum $P^\mu=(P^0,0,0,P^z)$ (and polarization $\epsilon$ for vector mesons). The nonlocal operator $O(\bar{\Gamma},\xi^-)$ is defined as
\begin{align}
O(\bar{\Gamma},\xi^-)=\bar\psi(\xi^-)\bar{\Gamma}\lambda^a W(\xi^-,0)\psi(0)
\end{align}
where $\bar{\Gamma}=\gamma^+\gamma_5$, $\gamma^+$, $\gamma^+\gamma_\perp$ correspond to pseudoscalar, longitudinally polarized vector, and transversely polarized vector meson LCDAs;
$\lambda$ is a Gell-Mann matrix as a flavor space projection, e.g. for pseudoscalar meson, $\lambda^a=\lambda^3$, $(\lambda^4\pm i\lambda^5)/2$, and $\lambda^8$ correspond to $\pi^0$, $K^\pm$, and $\eta$ meson states, respectively; the Wilson line $W(\xi^-,0)=P\exp [-ig_s\int^{\xi^-}_0 A^+(\eta^-)d\eta^-]$ is introduced to maintain the gauge invariance of the operator. 
The meson LCDA $\phi(\bar{\Gamma},x,\mu)$ is then defined through
\begin{align}\label{eq:LCDA}
{\cal F}(\bar{\Gamma},x,\mu)={\cal V}(\bar{\Gamma},\mu)\,\phi(\bar{\Gamma},x,\mu)
\end{align}
where
\begin{align}
{\cal V}(\bar{\Gamma},\mu)=\langle P,\epsilon|O(\bar{\Gamma},0)|0\rangle
\end{align}
is the renormalized matrix element of the local operator $O(\bar{\Gamma},0)$ which defines the decay constant~\cite{Ball:1998sk,Lu:2018obb}
\begin{align}\label{eq:decay_constant}
\langle P|O(\gamma^\mu\gamma_5,0)|0\rangle&=i f_P P^\mu\,,\\
\langle P,\epsilon_\parallel|O(\gamma^\mu,0)|0\rangle&=f_V^\parallel M_V\epsilon^{*\mu}_\parallel\,,\\
\langle P,\epsilon_\perp|O(\sigma^{\mu\nu},0)|0\rangle&=i f_V^\perp (\epsilon_\perp^{*\mu} P^\nu - \epsilon_\perp^{*\nu} P^\mu)\,
\end{align}
where $M_V$ is mass of the vector meson. In QCD, $f_P $ and  $f_V^\parallel$ are not renormalized while  $f_V^\perp $ depends on the scale. This definition guarantees the normalization of LCDA to be one,
\begin{align} \label{eq:LCDAnorm}
\int_0^1 dx\ \phi(\bar{\Gamma},x,\mu)=1\,.
\end{align}

To access the $x$-dependence of meson LCDA in LaMET, we consider Fourier transformation of an equal-time correlator $\langle P,\epsilon|\widetilde O(\Gamma,z)|0\rangle$ with spatial separation in $z$ direction
\begin{align}\label{eq:quasiDA_numerator}
\widetilde{\cal F}(\Gamma,x,P^z,\widetilde{\mu})=P^z\int \frac{dz}{2\pi}e^{i xz P^z}\langle P,\epsilon|\widetilde O(\Gamma,z)|0\rangle
\end{align}
where $\widetilde{\mu}$ is the renormalization scale for quasi-DA, 
\begin{align}\label{eq:tildeO}
\widetilde O(\Gamma,z)=\bar\psi(z)\Gamma \lambda^a  W(z,0)\psi(0)
\end{align}
is separated in the $z$ direction at equal time, and the Wilson line $W(z,0)=P\exp [-ig_s\int^{z}_0 A^z(z')dz']$. We define a quasi-DA through
\begin{align}\label{eq:quasiDA}
\widetilde{\cal F}(\Gamma,x,P^z,\widetilde{\mu})={\cal V}(\Gamma,\widetilde\mu)\,\widetilde\phi(\Gamma,x,P^z,\widetilde{\mu}).
\end{align}
Unlike the LCDA, the quasi-DA has support $x\in(-\infty,\infty)$. Note that $\widetilde O(\Gamma,0)=O(\Gamma,0)$, so the quasi-DA is also normalized to one
\begin{align}\label{eq:quasiDA_norm}
\int_{-\infty}^\infty dx\, \widetilde{\phi}(\Gamma,x,P^z,\widetilde{\mu})=1.
\end{align}
In order to avoid operator mixing \cite{Constantinou:2017sej,Chen:2017mzz,Chen:2017mie} on the lattice, we choose $\Gamma=\gamma^z\gamma_5$, $\gamma^t$, $\gamma^z\gamma_\perp$ for pseudoscalar, longitudinally polarized vector, and transversely polarized vector meson quasi-DAs, respectively. 

According to LaMET~\cite{Ji:2013dva,Ji:2014gla}, the distribution $\cal F$ and $\widetilde{\cal F}$ are related through a factorization formula, which can be derived using the method in Ref.~\cite{Izubuchi:2018srq}
\begin{align}\label{eq:factorization}
\widetilde{\cal F}(\Gamma,x,P^z,\widetilde{\mu}) =& \int_0^1 dy\, \widetilde C_{\Gamma}\left(x,y,{\tilde{\mu}\over\mu},{P^z\over \mu}\right) {\cal F}(\bar{\Gamma},y,\mu)\nonumber\\
& +\mathcal{O}\left(\frac{M^2}{(P^z)^2},\frac{\Lambda_{\rm QCD}^2}{(P^z)^2}\right)
\end{align}
where $\mathcal{O}\left(M^2/(P^z)^2,\Lambda_{\rm QCD}^2/(P^z)^2\right)$ are mass and higher-twist corrections. Since the choice of $\Gamma$ corresponds to a unique $\bar{\Gamma}$, we suppress the label $\bar{\Gamma}$ of the matching coefficient $\widetilde C_{\Gamma}$.
On the other hand, the renormalized local operators in Eqs. (\ref{eq:LCDA}) and (\ref{eq:quasiDA}) are related by
\begin{align}\label{eq:local_operator_relation}
{\cal V}(\bar{\Gamma},\mu)=\widetilde Z(\bar{\Gamma},\Gamma,\mu,\widetilde\mu){\cal V}(\Gamma,\widetilde\mu)
\end{align}
where $\widetilde Z(\bar{\Gamma},\Gamma,\mu,\widetilde\mu)$ contains kinematic factors in Eq. (\ref{eq:decay_constant}) and the scheme conversion factor when LCDA and quasi-DA are renormalized in different schemes.
Combining Eqs. (\ref{eq:factorization}) and (\ref{eq:local_operator_relation}), we have the matching formula between quasi-DA and LCDA~\cite{Ji:2015qla,Xu:2018mpf}
\begin{align}\label{eq:matching}
\widetilde\phi(\Gamma,x,P^z,\widetilde{\mu}) =& \int_0^1 dy\, C_{\Gamma}\left(x,y,{\tilde{\mu}\over\mu},{P^z\over \mu}\right) \phi(\bar{\Gamma},y,\mu)\nonumber\\
&+\mathcal{O}\left(\frac{M^2}{(P^z)^2},\frac{\Lambda_{\rm QCD}^2}{(P^z)^2}\right)
\end{align}
where $C_{\Gamma}=\widetilde Z\,\widetilde C_{\Gamma}$ is still perturbatively calculable.

\section{Renormalization}\label{sec:renormalization}
For each value of $z$, the RI/MOM renormalization factor $Z$ is calculated nonperturbatively on the lattice by imposing the condition that the quantum corrections of the correlator in an off-shell quark state vanish at scales $\{\widetilde{\mu}\}$~\cite{Constantinou:2017sej,Stewart:2017tvs}
\begin{align}
Z(\Gamma,z,a,\{\widetilde{\mu}\})=\left.\frac{\langle p'|\widetilde O(\Gamma,z,a)|p''\rangle}{\langle p'|\widetilde O(\Gamma,z,a)|p''\rangle_{\rm tree}}\right|_{\{\widetilde{\mu}\}}
\end{align}
where $(p')^2=(p'')^2$ in usual lattice setup; $\widetilde O(\Gamma,z,a)$ is the discretized version of $\widetilde O(\Gamma,z)$ on the lattice in Eq.~(\ref{eq:tildeO}) with spacing $a$; the bare matrix element $\langle p'|\widetilde O(\Gamma,z,a)|p''\rangle$ is obtained from the amputated Green's function $G(\Gamma,z,a,p',p'')$ of $\widetilde O(\Gamma,z,a)$, which is calculated on lattice, with a projection operator ${\cal P}$ for the Dirac matrix,
\begin{equation}\label{eq:amputated}
\langle p'|\widetilde O(\Gamma,z,a)|p''\rangle= \mbox{Tr}\left[G(\Gamma,z,a,p',p''){\cal P}\right].
\end{equation}
The UV divergence of the quasi-DA only depends on the operator $\widetilde O(\Gamma,z)$ itself, not the external states. In a higher-order Feynman diagram, it originates from the limit of all loop momentum components going to infinity, and is universal for all kinds of external states.
Therefore, we can choose a symmetric RI/MOM scheme:
\begin{align}\label{eq:Z}
Z_s(\Gamma,z,a,\mu_R,p^z_R)=\left.\frac{\langle p|\widetilde O(\Gamma,z,a)|p\rangle}{\langle p|\widetilde O(\Gamma,z,a)|p\rangle_{\rm tree}}\right|_{\{\widetilde{\mu}\}}
\end{align}
where $\{\widetilde{\mu}\}=\{p^2=-\mu_R^2,p^z=p^z_R\}$ and the dependence on $p^z_R$ is due to the breaking of Lorentz symmetry in the $z$ direction; ``symmetric" refers to setting the initial and final quark states to be the same, i.e. $|p'\rangle=|p''\rangle=|p\rangle$. This choice is the same as the renormalization factor for the quasi-PDF~\cite{Stewart:2017tvs,Liu:2018uuj}.

In a systematic calculation of LCDA, one starts with the bare correlator for the meson on the lattice,
\begin{align}
\widetilde{h}(\Gamma,z,P^z,a)=\langle P,\epsilon|\widetilde O(\Gamma,z,a)|0\rangle
\end{align}
which is renormalized and taken the continuum limit as
\begin{align}
&\widetilde{h}_R(\Gamma,z,P^z,\mu_R,p^z_R)\nonumber\\
&=\lim_{a\to 0} Z^{-1}_s(\Gamma,z,a,\mu_R,p^z_R)\widetilde{h}(\Gamma,z,P^z,a)\, ,
\end{align}
which is to be Fourier transformed into the $x$-space to obtain the distribution $\widetilde{\cal F}$ 
\begin{align}
&\widetilde{\cal F}(\Gamma,x,P^z,\mu_R,p^z_R)\nonumber\\
&=P^z \int \frac{dz}{2\pi}e^{i xz P^z}\widetilde{h}_R(\Gamma,z,P^z,\mu_R,p^z_R)\, .
\end{align}
${\cal V}(\Gamma,\mu_R)$ is given by $\widetilde{h}_R$ at $z=0$,
\begin{align}
{\cal V}(\Gamma,\mu_R)=\widetilde{h}_R(\Gamma,z=0,P^z,\mu_R,p_R^z)\, ,
\end{align}
which is frame independent and only depends on $\mu_R$. With $\widetilde{\cal F}$ and $\cal V$ calculated on the lattice, we acquire the quasi-DA
\begin{align}
&\widetilde{\phi}(\Gamma,x,P^z,\mu_R,p^z_R)\nonumber\\
&=P^z \int \frac{dz}{2\pi}e^{i xz P^z}\frac{\widetilde{h}_R(\Gamma,z,P^z,\mu_R,p^z_R)}{\widetilde{h}_R(\Gamma,z=0,\mu_R)}\,,
\end{align}
which satisfies the normalization condition in Eq.~(\ref{eq:quasiDA_norm}). Finally, we match quasi-DA in RI/MOM scheme to the LCDA in $\overline{\rm MS}$ scheme according to Eq.~(\ref{eq:matching}).

Since $\widetilde{\phi}(\Gamma,x,P^z,\mu_R,p_R^z)$ is independent of the UV regulator, we can calculate the matching coefficient in the continuum with perturbation theory using dimensional regularization. The one-loop result is provided in Sec.~\ref{sec:oneloop}.

\section{One loop matching coefficient}\label{sec:oneloop}
When the meson momentum $P^z$ is much greater than mass of the meson and $\Lambda_{\rm QCD}$, the quasi-DA in the RI/MOM scheme can be matched to LCDA through the factorization formula~\cite{Stewart:2017tvs,Izubuchi:2018srq,Ji:2015qla,Xu:2018mpf},
\begin{align}\label{eq:matching2}
&\widetilde\phi(\Gamma,x,P^z,\mu_R,p^z_R)\nonumber\\
=&\int_{0}^1 dy\, C_\Gamma\left(x,y,r,\frac{P^z}{\mu},{P^z\over p_R^z}\right) \phi(\Gamma,y,\mu)\nonumber\\
&+\mathcal{O}\left(\frac{M^2}{(P^z)^2},\frac{\Lambda_{\rm QCD}^2}{(P^z)^2}\right)\,,
\end{align}
where $r=\mu_R^2/(p^z_R)^2$. Note that the dependence of $C_\Gamma$ on $x$ and $y$ is different from the quasi-PDF case, which can be proved  using the same method in Ref.~\cite{Izubuchi:2018srq}. To obtain the matching coefficient from quasi-DA in RI/MOM scheme to LCDA in $\overline{\rm MS}$ scheme, we calculate their off-shell quark matrix element in perturbation theory by replacing the meson state $\langle P,\epsilon|$ in Eqs. (\ref{eq:LCDA}) and (\ref{eq:quasiDA}) to the lowest Fock state $\langle \bar Q(y P) Q((1-y)P)|$ where $y P$ and $(1-y)P$ are the momenta of the quark $Q$ and anti-quark $\bar Q$, respectively.

At tree level, the LCDA and quasi-DA with quark external state are 
\begin{align}\label{eq:tree_level}
\phi^{(0)}(\Gamma,x,y)=\widetilde{\phi}^{(0)}(\Gamma,x,y)=\delta(x-y).
\end{align}

In order to combine the ``real" and ``virtual" contributions (defined in Ref.~\cite{Stewart:2017tvs}) in a compact form at one-loop level, we introduce a plus function $\left[h(x,y)\right]_{+(y)}$ which is defined as 
\begin{align}\label{eq:plus}
\int dx [h(x,y)]_{+(y)} g(x)&=\int dx\ h(x,y)[g(x)-g(y)]
\end{align}
with two arbitrary functions $h(x,y)$ and $g(x)$. Following the procedure in Ref.~\cite{Stewart:2017tvs,Liu:2018uuj}, we need to take the on-shell ($P^2\to 0$) and large-momentum ($P^t\to P^z$) limits of the bare quasi-DA $\widetilde\phi_B(\Gamma,x,y,P^z,-P^2)$ to match it onto LCDA.  We obtain the bare matching coefficient
\begin{align}
&C^{(1)}_B\left(\Gamma,x,y,\frac{P^z}{\mu}\right)\nonumber\\
&=\widetilde\phi^{(1)}_{B}(\Gamma,x,y,P^z,-P^2)-\phi^{(1)}(\Gamma,x,y,\mu,-P^2)
\end{align}
where the subscript $B$ denotes ``bare"; $-P^2$ is the infrared (IR) divergence regulator, which is canceled in $C^{(1)}_B$ as expected. 
As we have showed in Eq. (\ref{eq:matching}), the matching coefficient from quasi-DA in RI/MOM scheme to LCDA in $\overline{\rm MS}$ scheme contains the matching factor of the factorization formula Eq. (\ref{eq:factorization}) as well as a perturbative conversion factor $\widetilde Z(\bar{\Gamma},\Gamma,\mu,\widetilde\mu)$. The combination of these two factors not only guarantee $C_\Gamma\left(x,y,r,P^z/\mu,P^z/p_R^z\right)$ to be unity after integaration over $x$, but also allow us to write $C_B(\Gamma,x,y,P^z/\mu)$ as a plus function by only considering the real contributions of the Feynman diagrams, even for the case of non-conserved current $\Gamma=\gamma^z\gamma_\perp$.

Since one needs to take the on-shell limit to obtain the bare matching coefficient, $C^{(1)}_B$, it is independent of the choices of gauge and projection operator $\cal P$ defined in Eq. (\ref{eq:amputated}). The results of the bare matching coefficients are
\begin{widetext}
\begin{align}\label{eq:bare_matching}
C^{(1)}_B\left(\Gamma,x,y,\frac{P_z}{\mu}\right)=\frac{\alpha_s C_F}{2\pi}\left\{
\begin{array}{lc}
\left[H_1(\Gamma,x,y)\right]_{+(y)}				& x<0<y\\
\left[H_2(\Gamma,x,y,P^z/\mu)\right]_{+(y)}		& 0<x<y\\
\left[H_2(\Gamma,1-x,1-y,P^z/\mu)\right]_{+(y)}	& y<x<1\\
\left[H_1(\Gamma,1-x,1-y)\right]_{+(y)}			& y<1<x
\end{array}\right.
\end{align}
where
\begin{align}
H_1(\Gamma,x,y)&=\left\{
\begin{array}{ll}
\frac{1+x-y}{y-x}\frac{1-x}{1-y}\ln\frac{y-x}{1-x}+\frac{1+y-x}{y-x}\frac{x}{y}\ln\frac{y-x}{-x} & \Gamma=\gamma^z\gamma_5{\;\rm and\;}\gamma^t\\
\frac{1}{y-x}\frac{1-x}{1-y}\ln\frac{y-x}{1-x}+\frac{1}{y-x}\frac{x}{y}\ln\frac{y-x}{-x} & \Gamma=\gamma^z\gamma_\perp
\end{array} \right.\, ,\\
H_2\left(\Gamma,x,y,\frac{P_z}{\mu}\right)&=\left\{
\begin{array}{ll}
\frac{1+y-x}{y-x}\frac{x}{y}\ln\frac{4x(y-x)(P^z)^2}{\mu^2}+\frac{1+x-y}{y-x}\left(\frac{1-x}{1-y}\ln\frac{y-x}{1-x}-\frac{x}{y}\right) & \Gamma=\gamma^z\gamma_5\\
\frac{1+y-x}{y-x}\frac{x}{y}\left(\ln\frac{4x(y-x)(P^z)^2}{\mu^2}-1\right)+\frac{1+x-y}{y-x}\frac{1-x}{1-y}\ln\frac{y-x}{1-x} & \Gamma=\gamma^t\\
\frac{1}{y-x}\frac{x}{y}\ln\frac{4x(y-x)(P^z)^2}{\mu^2}+\frac{1}{y-x}\left(\frac{1-x}{1-y}\ln\frac{y-x}{1-x}-\frac{x}{y}\right) & \Gamma=\gamma^z\gamma_\perp
\end{array} \right. \, .
\end{align}

Next we need to determine the counterterm of the quasi-DA in RI/MOM scheme. As we argued in Sec. (\ref{sec:renormalization}), we can use the renormalization factor for the quasi-PDF to renormalize the quais-DA, which leads to the RI/MOM counter-term
\begin{align}\label{eq:counterterm}
C^{(1)}_{CT}\left(\Gamma,x,y,r,\frac{P^z}{p^z_R}\right)=\left|\frac{P^z}{p^z_R}\right|\widetilde{q}^{(1)}\left(\Gamma,\frac{P^z}{p^z_R}(x-y)+1,r\right)_{+(y)}\,.
\end{align} 
$\widetilde{q}^{(1)}(\Gamma,x,r)$ is the real contribution of quasi-PDF at the RI/MOM subtraction scales $\mu_R$ and $p^z_R$. We choose Landau gauge, which is convenient  for lattice renormalization, and the minimal projection defined in Ref.~\cite{Liu:2018uuj} to calculate $\widetilde{q}^{(1)}(\Gamma,x,r)$. The results of $\widetilde{q}^{(1)}(\Gamma,x,r)$ for different spin structures are \cite{Liu:2018uuj,Liu:2018hxv},
\begin{align}
\widetilde{q}^{(1)}(\gamma^z\gamma_5,x,r)=\frac{\alpha_s C_F}{2\pi}\left\{
\begin{array}{lc}
\frac{3r-(1-2x)^2}{2(r-1)(1-x)}-\frac{4x^2(2-3r+2x+4rx-12x^2+8x^3)}{(r-1)(r-4x+4x^2)^2}+\frac{2-3r+2x^2}{(r-1)^{3/2}(x-1)}\tan^{-1}\frac{\sqrt{r-1}}{2x-1} & x>1\\
\frac{1-3r+4x^2}{2(r-1)(1-x)}+\frac{-2+3r-2x^2}{(r-1)^{3/2}(1-x)}\tan^{-1}\sqrt{r-1} & 0<x<1\\
-\frac{3r-(1-2x)^2}{2(r-1)(1-x)}+\frac{4x^2(2-3r+2x+4rx-12x^2+8x^3)}{(r-1)(r-4x+4x^2)^2}-\frac{2-3r+2x^2}{(r-1)^{3/2}(x-1)}\tan^{-1}\frac{\sqrt{r-1}}{2x-1} & x<0
\end{array} \right. ,
\end{align}
\begin{align}
\widetilde{q}^{(1)}(\gamma^t,x,r)=\frac{\alpha_s C_F}{2\pi}\left\{
\begin{array}{lc}
\frac{-3r^2+13rx-8x^2-10rx^2+8x^3}{2(r-1)(x-1)(r-4x+4x^2)}+\frac{-3r+8x-rx-4x^2}{2(r-1)^{3/2}(x-1)}\tan^{-1}\frac{\sqrt{r-1}}{2x-1} & x>1\\
\frac{-3r+7x-4x^2}{2(r-1)(1-x)}+\frac{3r-8x+rx+4x^2}{2(r-1)^{3/2}(1-x)}\tan^{-1}\sqrt{r-1} & 0<x<1\\
-\frac{-3r^2+13rx-8x^2-10rx^2+8x^3}{2(r-1)(x-1)(r-4x+4x^2)}-\frac{-3r+8x-rx-4x^2}{2(r-1)^{3/2}(x-1)}\tan^{-1}\frac{\sqrt{r-1}}{2x-1} & x<0
\end{array} \right. ,
\end{align}
\begin{align}
\widetilde{q}^{(1)}(\gamma^z\gamma_\perp,x,r)=\frac{\alpha_s C_F}{2\pi}\left\{
\begin{array}{lc}
\frac{3}{2(1-x)}+\frac{r-2x}{(r-1)(r-4x+4x^2)}+\frac{r-2x+rx}{(r-1)^{3/2}(1-x)}\tan^{-1}\frac{\sqrt{r-1}}{2x-1} & x>1\\
\frac{1-3r+2x}{2(r-1)(1-x)}+\frac{r-2x+rx}{(r-1)^{3/2}(1-x)}\tan^{-1}\sqrt{r-1} & 0<x<1\\
-\frac{3}{2(1-x)}-\frac{r-2x}{(r-1)(r-4x+4x^2)}-\frac{r-2x+rx}{(r-1)^{3/2}(1-x)}\tan^{-1}\frac{\sqrt{r-1}}{2x-1} & x<0
\end{array} \right. .
\end{align}

Finally, combining Eqs. (\ref{eq:bare_matching}) and (\ref{eq:counterterm}), we obtain the one-loop matching coefficient $C_\Gamma$ in Eq. (\ref{eq:matching2}),
\begin{align}
C_\Gamma\left(x,y,r,\frac{P^z}{\mu},\frac{P^z}{p^z_R}\right)=\delta(x-y)+C^{(1)}_B\left(\Gamma,x,y,\frac{P_z}{\mu}\right)-C^{(1)}_{CT}\left(\Gamma,x,y,r,\frac{P^z}{p^z_R}\right)+\mathcal{O}(\alpha_s^2).
\end{align}

\end{widetext}

\section{summary}\label{sec:sum}

In this work, we have pointed out that the quasi-DA can be renormalized in the RI/MOM scheme with the same renormalization factor that has already been calculated for the quasi-PDF case.
We have derived the one-loop matching coefficient that matches RI/MOM quasi-DA in the Landau gauge to $\overline{\rm MS}$ LCDA within the framework of LaMET. Our results include the matching coefficients for pseudoscalar, longitudinally polarized vector, and transversely polarized vector DAs with $\Gamma=\gamma^z\gamma_5$, $\gamma^t$, and $\gamma^z\gamma_\perp$, respectively. 
Our results are ready to be applied to extract the LCDAs from the lattice matrix elements of quasi-DAs.

\section*{Acknowledgments}
We are grateful to Yizhuang Liu, Xiangdong Ji, Yi-Bo Yang, and Jian-Hui Zhang for inspiring discussions. 
YSL is supported by Science and Technology Commission of Shanghai Municipality (Grant No.16DZ2260200) and National Natural Science Foundation of China (Grant No.11655002).
WW, JX, and SZ are supported in part by National Natural Science Foundation of China under Grant No.11575110, 11655002, 11735010 by Natural Science Foundation of Shanghai under Grant No.~15DZ2272100 and No.~15ZR1423100, Shanghai Key Laboratory for Particle Physics and Cosmology, and  by  MOE  Key Laboratory for Particle Physics, Astrophysics and Cosmology.
QAZ is supported by National Natural Science Foundation of China under Grant No.11621131001 and 11521505.
YZ is supported by the U.S. Department of Energy, Office of Science, Office of Nuclear Physics, from DE-SC0011090 and within the framework of the TMD Topical Collaboration. YZ is also grateful for the hospitality of Shanghai Jiao-Tong University during this stay at SKLPPC and T-D Lee Institute where this work was made possible.


\end{document}